%% file: bare_conf.tex
\documentclass[conference]{IEEEtran}
\ifCLASSINFOpdf
\else
\fi
\usepackage{amsmath}
\usepackage{mathrsfs}
\usepackage{graphicx}
\usepackage{url}
\usepackage{color, colortbl}
\usepackage{epsfig}
\usepackage{caption}

\DeclareMathOperator*{\argmin}{arg\,min}
\DeclareMathOperator*{\argmax}{arg\,max}

\newtheorem{example}{Example}

\newcommand{\Red}[1]{\textcolor[rgb]{1.00,0.00,0.00}{#1}}
\newcommand{\Blue}[1]{\textcolor[rgb]{0.00,1.00,0.00}{#1}}

\newcommand{\zy}[1]{\Blue{#1}}
\newcommand{\yh}[1]{\Red{#1}}
\newcommand{\tabincell}[2]{\begin{tabular}{@{}#1@{}}#2\end{tabular}}

\newcommand{\nop}[1]{}
\begin{document}
%
\title{Entity Suggestion by Example using a Conceptual Taxonomy}




%
\author{\IEEEauthorblockN{Yi Zhang\IEEEauthorrefmark{1},
Yanghua Xiao\IEEEauthorrefmark{2},
Seung-won Hwang\IEEEauthorrefmark{3}, 
Haixun Wang\IEEEauthorrefmark{4},
X. Sean Wang\IEEEauthorrefmark{5}, and
Wei Wang\IEEEauthorrefmark{6}
}
\IEEEauthorblockA{\IEEEauthorrefmark{1} \IEEEauthorrefmark{2} \IEEEauthorrefmark{5} \IEEEauthorrefmark{6} Fudan University, Shanghai, China,\\ \IEEEauthorrefmark{1} z\_yi11@fudan.edu.cn \\ \IEEEauthorrefmark{2} shawyh@fudan.edu.cn \\ \IEEEauthorrefmark{5} xywangcs@fudan.edu.cn \\ \IEEEauthorrefmark{6} weiwang1@fudan.edu.cn}
\IEEEauthorblockA{\IEEEauthorrefmark{3} Yonsei University, Seoul, South Korea\\ \IEEEauthorrefmark{3} seungwonh@yonsei.ac.kr}
\IEEEauthorblockA{\IEEEauthorrefmark{4} Facebook, Menlo Park, CA, USA \\ \IEEEauthorrefmark{4} haixun@gmail.com}
}


\maketitle

\begin{abstract}
\input{abstract}
\end{abstract}


%
\IEEEpeerreviewmaketitle

\input{Introduction}
  \label{sec:intro}

\input{RelatedWork}

\input{BackgroundAndBaseline}

\input{ProblemModel}

\input{ModelComputation}

\input{Experiment}

\section{Conclusion}\label{sec:con}
This paper studies entity suggestion by example, a technique with diverse applications.
Though many solutions exist, they are built mostly on the idea of co-occurrence in the text or web lists and exhibit limited effectiveness. In contrast, we leverage the new opportunities brought by many web-scale conceptual taxonomies with rich isA concept-instance relationships.
Specifically, we have proposed two probabilistic approaches, the first leveraging the typicality of concepts and entities to make the inference biased toward the more promising entities, and the second using an optimization solution to minimize the difference before and after the acceptance of a candidate entity. With these two models, we have solved the challenging problems of how to aggregate the conceptual information of the example entities by a Naive Bayes Model and a Noisy-Or model, how to find fine-grained concepts by an entity-based approach and a hierarchy-based approach and how to avoid the false positive in the inference due to the absence of contextual information. We have validated the effectiveness of our approaches using extensive evaluations with real-life data.

\bibliographystyle{IEEEtran}
\normalsize
\bibliography{reference}


\end{document}

%% file: abstract.tex
\nop{Entity suggestion by examples has been widely investigated in diverse scenarios. In a typical scenario, users provide a set of example entities as a query, and expect to get some entities completing the concept behind the given query. This problem has many applications such as related entity recommendation and query answering. However, existing solutions build on co-occurrence which is unaware of the concepts and their relationships of instances. Meanwhile, many web-scale conceptual taxonomies consisting of isA relationships between entity-concept pairs have enabled new opportunities to address this problem in a more systematic way. In this paper, we propose two probabilistic approaches to infer the entity that belongs to the concept implied by the exemplar entities. Our approaches, by leveraging the concept information of entities for the inference, significantly outperform existing ones, as validated by our extensive evaluations with real data.}
Entity suggestion by example (ESbE) refers to a type of entity acquisition query in which a user provides a set of example entities as the query and obtains in return some entities that best complete the concept underlying the given query. Such entity acquisition queries can be useful in many applications such as related-entity recommendation and query expansion. A number of ESbE query processing solutions exist in the literature. However, they mostly build only on the idea of entity co-occurrences either in text or web lists, without taking advantage of the existence of many web-scale conceptual taxonomies that consist of hierarchical isA relationships between entity-concept pairs. This paper provides a query processing method based on the relevance models between entity sets and concepts. These relevance models can be used to obtain the fine-grained concepts implied by the query entity set, and the entities that belong to a given concept, thereby providing the entity suggestions. Extensive evaluations with real data sets show that the accuracy of the queries processed with this new method is significantly higher than that of existing solutions.

%% file: Introduction.tex
\section{Introduction}

{\it Entity suggestion by example (ESbE)} has been widely investigated in different scenarios.
In a typical scenario,  a system accepts a set of example entities provided by a user as a query $q$, 
and retrieves a set of entities such that these entities, along with $q$, complete some concepts.
For example, a user may type \textsl{\{China, India, Brazil}\} as a query. It is quite possible that the user bears the concept \textsl{BRIC} in mind but cannot recall all of its members, so he/she enters these example entities of the concept for the purpose of retrieving the remaining ones. Then, the remaining entity \textsl{Russia} should be returned as the result. We give more examples in Table~\ref{ref:tb1}. Entity suggestion by example is also known as {\it entity list completion}~\cite{TREC, understanding}, {\it entity retrieval}~\cite{er2012,entity2014}, {\it entity recommendation}~\cite{YuX} or {\it entity query by example}~\cite{balog,Example} in different settings.

In general, the example entities can also imply some other concepts such as \textsl{country} in our previous example. However, a general concept usually leads to many less related entities. For example the concept \textsl(country) might mislead to the less related entities such as \textsl{US}. Hence, in ESbE, we aim to find fine-grained (specific) underlying concepts so that we can suggest most semantically related entities.
We will elaborate the selection of concepts in Section~\ref{sec:comp}.

ESbE has many applications. For example, in a search engine, a good recommendation of {\it 'related entities'}~\cite{2009} can be useful in encouraging users to click on the links to these entities~\cite{entityranking}. As another example, in a spreadsheet application, a tool that can automatically populate a spreadsheet or list can save on manual labor. For instance, when a user wants to create a table of BRIC countries in the spread-sheet, it may reduce his/her effort significantly if the table can be automatically populated after a few examples are entered. 
In general, ESbE can be viewed as a type of query by example~\cite{QBE}, and can be useful in some question answering systems~\cite{qa}. With more and more knowledge bases published, entity information becomes increasingly abundant and queries by exemplar entities on knowledge bases will become more popular~\cite{Exemplar, QAG}.

\nop{This problem has many applications.
First, {\it this problem can be viewed as a recommendation of {\it 'related entities'}}~\cite{2009}. A rational recommendation of related entities will encourage users to click the links to them~\cite{entityranking}. Second, {\it it can be used in the automatically populating spreadsheets or lists}.
Users may want to create a table of BRIC countries in spreadsheet, and automatic
population after a few examples would save human labors significantly.
Third, {\it the problem can be viewed as queries by examples}~\cite{QBE}, which is a typical query answering task~\cite{qa}. With more and more knowledge bases published, entity information becomes ever-increasing abundant. Queries by exemplar entities on knowledge graphs will become a popular query answering task~\cite{Exemplar, QAG}.}

\begin{table*}
\centering
\normalsize
\caption{Entities and Possible Fine-grained Underlying Concepts}
\begin{tabular}{|c|c|c|} \hline
Entity List&A Suggested Entity&A Possible Fine-grained Underlying Concept\\ \hline
China, India, Brazil&Russia&BRIC\\ \hline
Alibaba, Tecent&Baidu&BAT\\ \hline
swimming, marathon&bicycle ride&Ironman Triathlon\\ \hline
Islam, Buddhism&Christianity&The three major religions\\ \hline
Standard Poor's, Moody's&Fitch Group&Big Three(credit rating agency)\\ \hline
Roger Federer, Rafael Nadal, Novak Djokovic&Andy Murray&Big Four(tennis)\\ \hline\end{tabular}
\label{ref:tb1}
\end{table*}

\paragraph*{\bf Weakness of Previous Approaches} 

Many solutions to ESbE have been developed. These solutions can be classified into the following three categories.
The solutions in the first category tend to use co-occurrence as the basic recommendation mechanism. A well-known example is Google Set. The basic idea is to recommend the entities that most frequently co-occur with the example entities.
The solutions in the second category assume that the query set belongs to some lists and estimate how likely it is that each item belongs to a list containing the query items. An example in this category is SEISA~\cite{Y}.
The solutions in the third category rank all the entities based on how much their properties overlap with those of the example entities, and take the highest ranked entities not already in the query as the result.

\nop{The third line~\cite{Example} ranks all entities by their overlap with the properties of the example entities and adds the highest ranked entity not belonging to the target set.}

However, these solutions, being unaware of the concepts, especially fine-grained concepts underlying the example entities,  suffer from  the following limitations:
\begin{enumerate}
\item 
First, {\it co-occurrence does not necessarily imply conceptual coherence}~\cite{cc} {\it between the recommended entities and the example ones}. Best result entities are often those that share the same concepts with the example entities. For example, when a user searches with \textsl{\{China, India, Brazil\}}, the intention is very likely to find the \textsl{BRIC countries}. In general, \textsl{Russia} is a good entity to be recommended since entity \textsl{Russia} is better in keeping the concept coherent than entity \textsl{USA} is even if it frequently co-occurs with all the example entities respectively. Note that it is unrealistic to count the number of co-occurrence among a set of entities in advance by enumerating all possible combinations of entities even in an off-line procedure.  
\item 
Second, {\it the membership of entities in lists may be too weak a signal to be used to recommend the right entity.} In the state-of-the-art methods for set expansion SEAL~\cite{wang2007} and SEISA~\cite{Y}, web lists are used to construct an entity-list-membership network and candidate entities are ranked by a random walk based measure. In general, one should recommend an entity more strongly if it can be reached with higher probability by random walk starting from the given entities. However, because of the weak membership between entities and lists, the random walk along the membership edges is likely to find noisy entities. Even when there are lists containing all query examples, we cannot decide which is more desirable without knowing the concept behind it. They may refer to a concept too general to introduce false positives, or even they are lists randomly involving some entities from the web. For example, given \textsl{\{China, India, Brazil\}}, a very typical list they belong to may be a list of many countries. Then some other common entities in this kind of lists such as \textsl{USA, Britain} etc. may be ranked high in the results. These typical developed countries should obviously not be suggested first when given some of the BRIC countries.
\item
Third, {\it considering only the overlap of the properties of the example entities and those of the candidate entities does not always lead to the correct set expansion.} For example, given three entities \textsl{\{Industrial and Commercial Bank of China, China Construction Bank, Bank of China\}}, though the entity \textsl{China Merchants Bank} shares with most of the query entities in terms of its properties such as \textsl{location, owner} and so on, this entity may not be a good answer. The reason is because \textsl{Big Four State-owned Bank of China} may be the most relevant and fine-grained concept behind the three given banks while this concept does not cover \textsl{China Merchants Bank}. (The other Big-Four is \textsl{Agriculture Bank of China}.)

\nop{
\item
Third, {\it a single concept in general is not enough to capture the meaning of the example entities}~\cite{Y,R}. For example, for the input entities \textsl{\{Canon, Sony, Nikon\}}, neither  \textsl{camera brands} nor \textsl{Japanese companies} can individually characterize the concepts of the given entities.  Instead, a combination of these two concepts, or more generally, an aggregation of concept distributions is desired. Thus, the entity \textsl{Olympus} should be added into the list rather than \textsl{Kodak} because \textsl{Olympus} is both a Japanese company and a camera brand but \textsl{Kodak} is an American camera brand.

}
\end{enumerate}

Of course, entity suggestion can be quite subjective, and the best entities to recommend may vary from person to person.
Our work is to recommend entities by determining the fine-grained concepts underlying the query examples. Thus, we will give priority to the entities related to a fine-grained underlying concept which we think is a general sense for a generic user.  
With enough contextual information, our method may be extended to provide personalized entity suggestion, but this extension is beyond the scope of this paper.

\paragraph*{\bf Advantages of Conceptual Taxonomies}
Recently, many web-scale conceptual taxonomies consisting of isA relationships between entity-concept pairs, such as Microsoft's Probase  and Google's isA database, have become available.
These knowledge bases are constructed by Hearst patterns from a web corpus.
The abundant concept information in these knowledge bases brings us new opportunities to process ESbE queries. 
\begin{enumerate}
\item First, {\it the taxonomy allows us to explicitly and accurately model the concepts of example entities as well as their complicated relationships.} As we have seen in our examples above, the concept modeling is critical for ESbE. A large scale conceptual taxonomy allows us to explicitly represent the semantic of an entity or an entity list by its concept distributions.

\item Second, {\it the frequency information in taxonomy makes the inference more accurate}.
Some taxonomies, such as Probase, have the frequency information for the isA pairs. This allows the computation of many desired metrics, such as the typicality of an entity/concept. 
\nop{when observing a concept/entity available.} For example, \textsl{emerging market} is more typical than \textsl{country} as a concept for the example entities \textsl{ \{China, India, Brazil}\}. Using this more typical concept allows us to infer \textsl{Russia} instead of some other country entities.

\item Third, {\it the hierarchical structure of the conceptual taxonomies enables more accurate inference of the result entity.} Note that the hierarchy can be used to estimate the specificness of a concept. For example, \textsl{China} is\textsl{A} \textsl{country}, and also a \textsl{developing country}. Based on the taxonomies, we also know that a \textsl{developing country} is also a \textsl{country}. Thus, \textsl{developing country} is a relatively more specific concept. Using this concept allows to suggest more related entities by reducing the possibilities of false positives. 

\end{enumerate}

\paragraph*{\bf Challenges}
In this paper, we propose two probabilistic approaches to infer the entities that belong to the concepts implied by the exemplar entities.
Our approaches sufficiently utilize the concept information of entities for inference.
To do so, we face the following challenges:

First, {\it how to aggregate the conceptual information of example entities without introducing extra noise is still challenging.} To see this, we show that some naive aggregations, such as finding the {\em least common ancestor} (LCA) concept of query entities, fail in our running example. The LCA concept of \textsl{\{China, India, Brazil\}} in Probase consists of concepts such as \textsl{country, nation} and so on. Thus, it is very likely to return some entities such as \textsl{USA} by finding other entities of the LCA concepts such as \textsl{country}. To overcome this challenge, we propose a Bayesian approach and a Noisy-Or model to evaluate the relatedness between the concepts and the examples to represent the semantic distribution for the example entities in our solution.

Second, {\it how to determine fine-grained concepts instead of concepts too general from the concepts related to the query examples is difficult.} General concepts may introduce more false positive entities than fine-grained concepts may, thus we propose two kinds of methods to solve this problem. The first one is directly penalizing popular concepts which is an entity based approach, and the second one is a granularity-aware approach which is based on the hierarchical structure of the taxonomies.

Third, even if desired fine-grained concepts can be found, a false positive may still be generated without careful treatment. Since isA relationships between entities and concepts are extracted from a large corpus by patterns, some false positive entities may already exist in the list of the entities of a concept. For example, \textsl{China} is mentioned as a developed country in some corpus, thus it appears in the entity list of \textsl{developed country} as well. However, by a careful examination, one may notice that \textsl{China} is more closely related to the concept \textsl{developing country} in the taxonomy.
\nop{
entity pairs sharing the same concept may not be mentioned in the same entity list of the concept and also conflict concepts may share the same entities.  For example, \textsl{country} concept of 200+ entities of Probase is populated by 'countries such as X and Y' patterns, from which some entities such as \textsl{Mongolia} and \textsl{Canada} are rarely observed. However, it does not mean that \textsl{Mongolia} or \textsl{Canada} is not a country. Another example is that \textsl{China} appears in the both entity lists of \textsl{developed country} and \textsl{developing country} which does not mean that China is both a developed and developing country.
Second, {\it even though a desired concept exists, absence of contextual information complicates entity inference}.
Although Probase has isA relationships between entities and concepts, contextual information is lost when the isA relationships are extracted. As a result, entity pairs sharing the same concept may not be mentioned in the same list.
.
}
To address this challenge, we propose a probabilistic relevance model (in Section~\ref{subsec:1}), which leverages the typicality of entities to every related concept to make the inference bias towards the promising entities. We also model our problem as an optimization problem, which minimizes the difference before and after the admission of a candidate entity (in Section~\ref{subsec:2}). The optimization model leads us to promising entities.

The rest of the paper is organized as follows.  
In section~\ref{sec:2}, we review the related work. In section~\ref{sec:3}, we introduce some background knowledge and present the baseline method. In section~\ref{sec:pm}, we propose two probabilistic models for our problem. Section~\ref{sec:comp} elaborates how we compute the models. We present the experimental study in Section~\ref{sec:exp} and conclude the paper with Section~\ref{sec:con}.

%% file: RelatedWork.tex
\section{Related Work} \label{sec:2}
In this section, we review the related work, which can be classified into three categories: {\it related entity recommendation}, {\it entity set expansion} and {\it short text conceptualization}.

\subsection*{Entity Recommendation}
Related entity recommendation can be categorized into the following two categories:
First, to
{\it recommend related entities for search assistance},
Blanco et al.~\cite{Blanco} proposed a recommendation engine Spark to link a user's query  word to an entity within a knowledge base and recommend a ranked list of the related entities.
To guide user exploration of recommended entities, they also proposed
a series of features to characterize the relatedness between the query entity and the related entities.
Unlike this work assuming a single entity as a query, our work recommends related entities for a set of query entities of the same hidden concepts. Steffen et al.~\cite{aspect} proposed a similar entity search considering diversity, however, in our work, we assume an entity linked to related and fine-grained concepts more important.

Second, for
{\it query assistance for knowledge graphs},
GQBE~\cite{GQBE} and exemplar queries~\cite{Exemplar} studied how to retrieve entities from a knowledge base by specifying example entities. For example, the input entity pair \textsf{\small\{Jerry Yang, Yahoo!\}} would help retrieve answer pairs such as \textsf{\small\{Sergey Brin, Google\}}. Both of them projected the example entities onto the RDF knowledge graph to discover result entities as well as the relationships around them. GQBE used the weighted graph as the underlying model and exemplar queries used the edge-weighted one. All these works used the subgraph isomorphism as the basic matching scheme, which in general is costly. 
Our objective is to infer entities that preserve the semantic of the example entities. All the example entities generally share the same type.

\nop{ These two works are different with us because our query entities are parts of a conceptualized list instead of some entities with certain relationships, and our goal is to infer the entities left rather than finding another list with similar relationships in the list. Furthermore, there is no type information in Probase. 
}

\nop{survey the following two works: http://ranger.uta.edu/~cli/pubs/2014/gqbe-icde14demo-cameraready-nov13.pdf
http://event.cwi.nl/grades2014/11-jayaram.pdf}

\subsection*{Entity Set Expansion}

\nop{\zy{not clear!! Many commercial providers now support entity-oriented search, dealing with entity types such as people, companies, services and locations. These previous works on {\it Entity Retrieval} mostly focused on establishing associations between topics, documents, or among entities themselves. In Hema~\cite{H}'s work, it models an entity as a word distribution with summarization in order to do question answering.
}}

The goal of this line is, given a set of seed entities,
to discover other entities in the same concept.
Google Sets~\cite{Google} is a product implementation used
to populate a spreadsheet after users provide some instances as examples. 
Inspired by Google Sets, many research work followed~\cite{Bayesian,Y,R,L,P},
to measure the membership strength of an item for a hidden concept exemplified by query entities.  

However, this work assuming a single concept (always a too general concept) to explain the query set,
cannot disambiguate when the query set conceptualizes to multiple fine-grained concepts, such as a camera brand and Japanese company.
Our work assumes a query set can bear multiple fine-grained concepts, and aggregates a concept distribution to accurately infer related entities that reflect all the related concepts. 

Related problems include semantic search tasks studied in
Bron et al.~\cite{Example}, of taking example instances and the textual representation of a relation, to complete the list of examples.
Another example is harvesting tables on the Web, and retrieving the table that completes the example instances and description~\cite{Y,wang2014}.
Compared to this work, our task is more challenging relying solely on examples without explicit description of a relation or table. 

\nop{
Explicit Semantic Analysis(ESA)\cite{Egozi} is a method that represents the meaning of texts in a high-dimensional space of concepts derived from Wikipedia, and assesses the relatedness of texts in this space by comparing the corresponding vectors with conventional metrics. However, this method does not take several entities together into consideration as well. It does not mine the common concept which can not solve our problem.}

\subsection*{Text Conceptualization}

Conceptualization aims to map a short text to a set of concepts as a mechanism of understanding text.  Lee et al. ~\cite{Dongwoo}  proposed context-dependent conceptualization to capture the semantic relations between words by combining Latent Dirichlet Allocation with Probase. Song et al.~\cite{Song} developed a Bayesian inference mechanism to conceptualize words and short texts. 
The ultimate objective of conceptualization is to find the concepts that best capture the semantic of the short texts. 
However, conceptualization combines all related concepts together without considering the semantic granularity of the concepts which increases the risk of recommending false-positives. Our work is to find fine-grained concepts underlying the query, and the next step of entity inference, of identifying the most related entity, is beyond the scope of conceptualization as well.

%% file: BackgroundAndBaseline.tex
\section{Background and Baseline Approach} \label{sec:3}

In this section, we briefly review the conceptual taxonomy Probase upon which our solutions are built. We also argue that the straightforward solution based on Probase and the baseline solution based on the $k$-nearest neighbor algorithm cannot solve our problem. 

\subsection{Probase and isA relationships}

We use a web-scale conceptual taxonomy for the inference of the suggested entities.
Probase~\cite{wu} is a universal, general-purpose, probabilistic taxonomy automatically constructed from a corpus of 1.6 billion web pages. 
Probase contains 2.7 million concepts and 4.6 million isA (a.k.a., hypernym-hyponym) relationships among the concepts/entities, which is suitable for us to describe the query entities. 
Each isA relationship saying ($e$ isA $c$) is associated with the frequency ($n(e,c)$) that $e$ isA $c$ is observed from the corpus. The frequency allows us to compute the {\it typicality} of $e$ under concept $c$, i.e., $P(e|c)$, which can also be interpreted as the probability that $e$ is an instance of $c$.

Given Probase, a baseline of set expansion is to find the {\em least common ancestor} (LCA) concept of the query entities, then return an instance with the highest typicality for the LCA concept identified.
However, it does not perform well, as the LCA concept may not exist in Probase.
Furthermore, even the desired concept is identified, entity inference remains
as a challenge, as typicality does not reflect relationships with query
entities.

\subsection{A KNN based Approach}

As another baseline, we consider
a $k$-nearest neighbor (KNN) based solution
of returning the top-$k$ entities with the highest $sim(q, e)$.

Thus, the key is defining $sim$ using the concept distribution of entities
derived from Probase.
Specifically, given the query entity set $q =$\{$e_{1}$, $e_{2}$,..., $e_{n}$\}, each entity $e$ is associated with a concept vector $\mathbf{C}(e)=\{<c_{i}, P(e|c_i)>\}$, where $P(e|c_i)$ is the typicality of the instance $e$ given the concept $c_i$. Given Probase, $P(e|c_i)$ can be computed by the following equation:

\begin{equation}
\normalsize
P(e|c_i)=\frac{n(e, c_i)}{n(c_i)}
\label{eq:pec}
\end{equation}
where $n(c_i)$ is the number of occurrences of the concept $c_i$ in Probase, and $n(e,c_i)$ is the number of occurrences of $e$ as an instance of $c_i$.

Similarly, we can define the concept vector $\mathbf{C}(q)=\{<c_{i}, P(q|c_i)>\}$ for the query entity set $q$, such that each $c_i$ is a concept of at least one entity in $q$ and $P(q|c_i)$ is the typicality to observe any one entity in $q$ under the concept $c_i$ in Probase. $P(q|c_i)$ can be computed by the following equation:
$$P(q| c_i)=\frac{\sum_{e\in q}{n(e,c_i)}}{n(c_i)}$$

Once concept vectors for $q$ and $e$ are defined, similarity can be
computed as the cosine similarity with each other.
One obvious weakness of the KNN approach is that it is too costly. The time complexity is $O(|E_P||C_P|)$, where $|E_P|$ and $|C_P|$ are the numbers of entities and concepts in Probase respectively, because any entity in Probase is a candidate (overall $O(|E_P|)$ entities) and  the cosine similarity computation consumes $O(|C_P|)$ time. Note that Probase contains millions of entities and concepts. Hence, the complexity is generally unacceptable.

In the next section, we propose probabilistic models with higher accuracy and efficiency.

\nop{
A revised method is to filtering concepts who has low co-occurrence with the entity, but it will scarifies the performance.

A probabilistic method we will mention following has better performance than this basic method.
}

%% file: ProblemModel.tex
\section{Problem Model}\label{sec:pm}
In this section, we propose two models to solve our problem. 
The first model seeks to maximize a probabilistic based relevance ranking measure.
The second aims to minimize the difference between the concept distributions of query entities before and after its acceptance of the suggested entities. 

Table~\ref{tab:notation} lists the notations we use.

\begin{table}
\centering
\normalsize
\caption{Notation Table}
\begin{tabular}{|c|c|} \hline
Notation&Description\\ \hline
$E$&the universal set of entities\\ \hline
$q$&the set of query entities\\ \hline
$C$&the universal set of concepts\\ \hline
$rel(q,e)$& the relevance of an entity $e$ to $q$\\ \hline
$P(e|c_i)$& \tabincell{c}{the typicality of the entity $e$ \\given the concept $c_i$}\\ \hline
$P(c_i|e)$& \tabincell{c}{the typicality of the concept $c_i$ \\given the entity $e$}\\ \hline
$P(c_i|q)$& \tabincell{c}{the typicality of the concept $c_i$ \\given a query set $q$} \\ \hline
$P(c_i)$& the typicality of concept $c_i$\\ \hline
$\delta(c_i)$& the indicator of $c_i$ being a fine-grained concept\\ \hline
$c(e)$& the concept set of the entity $e$\\ \hline
$C_q^k$& \tabincell{c}{the set of $k$ fine-grained concepts \\underlying the query $q$}\\ \hline
$h(q_i|c)$& \tabincell{c}{the expected number of steps \\starting from $q_i$ to $c$}\\ \hline  
\end{tabular}
\label{tab:notation}
\end{table}

\subsection{A Probabilistic Relevance Model} \label{subsec:1}
When given a set of query entities $q=\{q_i|q_i \in E\}$, we model the relevance of an entity $e$ to $q$ with $rel(q,e)$, which can be interpreted as the likelihood that a real person will think of the entity $e$ when he/she observes the entities in the query $q$. Thus, our objective is to find the entity whose relevance is the highest:
\begin{equation}
\normalsize
\argmax_{e\in E-q} rel(q,e)
\label{eq:obj}
\end{equation}

Then, the key is to define the relevance function. 

Consider the psychological procedure of a user to infer an entity by formulating a set of example entities. A real user tends to formulate the query by referring to some concepts of the examples as well as the target entity. The concepts referred to describe the one or more aspects of these entities. For example, given  \textsl{\{China, India, Brazil\}}, two concepts \textsl{\{developing country, emerging market\}} naturally come to our mind. We define $P(c_i|q)$ to be the typicality that $c_i$ is referred to when we are presented with $q$. The most typical entity under the concepts the query referred to tends to be recommended. For the example above, in both of the two concepts, \textsl{Russia} is a typical entity. Let $P(e|c_i)$ be the typicality of $e$ given concept $c_i$, which can be computed by Eq.~\ref{eq:pec}. Thus, we have the following ranking function:
\begin{equation}
\normalsize
rel(q,e)=\sum_iP(e|c_i)P(c_i|q)
\end{equation}

Clearly, the ranking function is positively correlated to the two factors $P(e|c_i)$ and $P(c_i|q)$, which reflect the following two principles: 
(1) A typical instance should be recommended; (2) If a concept is more likely to be  associated with the query entities, its typical instance deserves to be recommended. The summation over all concepts reflects the fact that if there are many concepts leading to an instance $e$, it should be recommended. 

\paragraph*{ Interpretation of $P(c_i|q)$}

The direct interpretation of $P(c_i|q)$ is as follows. Suppose there exists an {\it ideal} concept for the query entities, which under some circumstances holds true. For example,  \textsl{\{China, India, Brazil\}} usually implies that the best concept is \textsl{BRIC} countries, which directly helps us find the appropriate entity \textsl{Russia}. Thus, $P(c_i|q)$ can also be interpreted as the typicality that $c_i$ is the \textsl{ideal} concept of $q$. 

Otherwise, the ideal concept does not always exist in the conceptual taxonomy. For example,
it is hard to give an explicit-yet-simple concept to describe \textsl{\{Baidu, Tecent, Alibaba\}}. They refer to the three biggest IT companies in China (BAT for short). 
In these cases,  $P(C|q)$ can be interpreted as the concept distribution of the query entity set.
For example, for the query entity set \textsl{\{China, India, Brazil\}}, \textsl{\{developing country, emerging market\}} are two representative concepts to describe the semantics of the query entity set. 
Thus, each concept $c_i$ can be aggregated with a weight $P(C=c_i|q)$ (for short $P(c_i|q)$), to represent the semantics of $q$ together.

However, some concepts are too general to be the ideal concepts to suggest other entities. In our running example, \textsl{country} is a concept which can summarize all the query examples but it makes little difference in evaluating other entities of countries. These concepts are relatively vague to characterize the given entities. Thus, we introduce an item $\delta(c_i)$ to help choose the ideal concepts given $q$. Then our new ranking function will be:

\begin{equation}
rel(q,e)=\sum_{i}P(e|c_i)P(c_i|q)\delta(c_i)
\end{equation}

We propose two strategies to compute $\delta(c_i)$, one is to penalize the popular concepts based on the entities of the concepts, the other one is to find fine-grained concepts based on the hierarchical structure of the conceptual taxonomy. We will elaborate them as well as the computation of $P(c_i|q)$ in Section~\ref{sec:comp}.

\subsection{A Relative Entropy Model}\label{subsec:2}

An alternative model is to use a concept distribution of query entities. We find the $e$ such that its admission into $q$ has the least impact on the original concept distribution. 
More formally, $P(C|q)$ is the concept distribution given query entity set $q$, which can be represented as a set of vectors $\{<c_i, P(c_i|q)>\}$. We just need to find the entity $e$ such that $P(C|q, e)$ is closest to $P(C|q)$. A popular measure of the distance between two probability distributions is KL-divergence, also known as relative entropy. Thus, our problem can be modelled as:
\begin{equation}
\argmin_{e\in E-q} KL(P(C|q),P(C|q,e))
\label{eq:kl}
\end{equation}
where KL-divergence is defined as:
\begin{equation}
KL(P(C|q), P(C|q,e))=\sum_{i=1}^n P(c_{i}|q)\times log(\frac{P(c_{i}|q)}{P(c_{i}|q, e)} )
\label{eq:kld}
\end{equation}

KL-divergence between $P(C|q)$ and $P(C|q, e)$ characterizes the expectation of the logarithmic difference between the two probability distributions. Note that KL-divergence is not symmetric, which means that $KL(P(C|q), P(C|q,e))\neq KL(P(C|q,e), P(C|q))$. The rationality of our optimization function is that the posterior typicality of $c_i$ observing $q$ (i.e. $P(c_i|q)$) is more confident than observing $q$ and an arbitrary entity $e$. Furthermore, as the same reason described in section~\ref{subsec:1}, Eq.~\ref{eq:kld} should also carefully choose the concepts by evaluating $\delta(c_i)$ and $P(c_i|q)$, which will be elaborated in the next section. 

Next, we justify why minimizing the distribution disparity in our problem 
setting is an effective objective.

\paragraph*{ Rationality of the objective function}

We justify the minimization objective by a hypothesis test. 
The basic idea is to show that the admission of a right entity will preserve the 
concept distribution,  with statistical significance, under the comparison to a null model.
Specifically, for each query entity set $q$ with the ground truth, we choose the answer entity $e$ to compute $d_1=KL(P(C|q),P(C|q, e))$, and randomly choose another entity $e'\neq e$ which shares at least one concept with any query entity in $q$ to compute $d_2=KL(P(C|q),P(C|q, e'))$. We repeat the random selection 50 times and summarize the average of $d_2$, denoted by $\bar{d}_2$.  Thus, for query entity set $q$, we have a pair of distances $(d_1, \bar{d}_2)$.
We calculate these paired distances for 50 different $q$. 

Then, we form the null hypothesis as: {\it there is no difference between  $d_1$ and  $\bar{d}_2$}. 
We test the hypothesis on all the pairs of distances by {\it paired t-test}.
The result shows that the $P$-value score is less than $0.001$, which is much smaller than the significance level $0.05$. 
The results suggest that there is sufficient statistical evidence to reject the null hypothesis. In other words, the admission of the answer entity leads to a smaller distribution distance than a randomly selected one.

\nop{ It examines whether there is a sufficient statistical evidence to support the hypothesis that the mean of the difference of every pair of distance equals zero, and it is defined as follows:
\begin{equation}
H_0: avg\{|d_{i1}-\bar{d}_{i2}|\}_{d_{i1}\in D_1, \bar{d}_{i2}\in \bar{D}_2}=0
\end{equation}
\begin{equation}
H_1: avg\{|d_{i1}-\bar{d}_{i2}|\}_{d_{i1}\in D_1, \bar{d}_{i2}\in \bar{D}_2}\neq 0
\end{equation}
where $H_0$ is the null hypothesis, $H_1$ is the alternative hypothesis. }

\begin{figure}
\centering
\epsfig{file=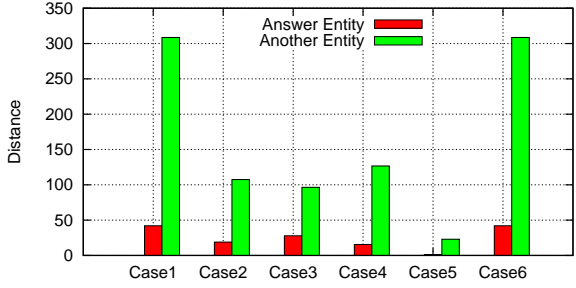,scale=0.9}
\caption{Case study on the objective function}
\label{fig:sixobf}
\end{figure}

 Figure~\ref{fig:sixobf} shows that our observation above consistently holds for six cases listed in Table~\ref{ref:tb1}.

%% file: ModelComputation.tex

\section{Model Computation}\label{sec:comp}

In this section, we elaborate how to compute $\delta(c_i)$ and $P(c_i|q)$ which are two common parts of the two probabilistic models that we have not discussed in detail. We propose two approaches for $\delta(c_i)$ and $P(c_i|q)$ respectively .
\nop{We first propose a Naive Bayes based approach, and then, propose a more efficient approach for the computation.}

\subsection{$\delta(c_i)$ Computation}
As we have mentioned in Section~\ref{sec:pm}, $\delta(c_i)$ evaluates the concepts which will be used as a latent variable to suggest entities. The basic idea is to find fine-grained underlying concepts for queries instead of those which are too general or too specific. A concept too general may be very likely to introduce noisy entities leading to low precision, while a concept too specific may be difficult to find related entities leading to low recall. Therefore, we propose two approaches here, the first one is to directly penalize popular concepts based on the entities of the concept, the second one is to use a granularity-aware model to find fine-grained concepts based on the hierarchical structure of the conceptual taxonomy .
   
\subsubsection{Penalizing Popular Concepts}
From the perspective of entities to observe concepts, a concept with more entities may be more general. For example, in Probase, \textsl{country} has 2648 entities, \textsl{developing country} has 149 entities, while \textsl{growing market} only has 18 entities. It is obvious that \textsl{country} is a concept more general than \textsl{developing country} and \textsl{growing market}, which is rational according to the institution of human beings. Thus, some concepts may have high $P(c_i|q)$ due to their popularity, such as
\textsl{company} for \textsl{\{Tencent, Baidu\}}. These concepts are relatively vague to characterize the given entities because they also contain many other entities and may even lead to the issue of semantic drift. It motivates us to introduce the typicality of the concept $P(c_i)$ to penalize a popular concept. 

Then, $\delta(c_i)$ will be:
\begin{equation}\delta(c_i)=\frac{1}{P(c_i)}
\label{eq:rank}
\end{equation} 

where $P(c_i)$ is defined as follows:
\begin{equation}
P(c_i)=\frac{n(c_i)}{\sum_{c\in C}{n(c)}}
\end{equation}
where $n(c_i)$ is the number of occurrence of $c_i$ derived from Probase data.

\subsubsection{Granularity-aware Approach}
When using $P(c_i)$ to penalize popular concepts, a potential problem may be bias to the concepts which are too specific which may lead to more difficulty in finding related entities since specific concepts always contain few entities. Thus, we propose a new method considering granularity of the concepts by using the measure of hitting time to the query based on the hierarchy of the concepts. We first give the computation method and then state the rationality.

Let $C_q^k$ be a set with $k$ fine-grained concepts underlying query $q$, the expected hitting time $H(q|C_q^k)$ of the random walk is the sum of the expected number of steps before each $q_i\in q$ is visiting a concept $c$ in $C_q^k$.

Thus, our target becomes finding a set $C_q^k$ such that:
\begin{equation}
\argmin_{C^k_q}{H(q|C^k_q)}
\end{equation}
Here $k$ can control the number of the concepts we choose, and the larger the $k$ is, the more general concepts will be introduced into the set, and to be used to suggest entities.

Naturally, $H(q|C_q^k)$ can be computed as follows:

\begin{equation}
H(q|C_q^k)=\sum_{q_i\in q}{\sum_{c\in C_q^k}{h(q_i|c)}}
\end{equation} 
where $h(q_i|c)$ means the expected number of steps before a query entity $q_i$ is visiting a concept $c$.

It can be easily verified that the hitting time satisfies the following system of linear equations~\cite{random}:

\begin{equation}
\left\{
\begin{aligned}
h(q_i|c)&=0,&if q_i=c\\
h(q_i|c)&=1+\sum_{c'\in c(q_i)}P(c'|q_i)h(c'|c),&if q_i\neq c
\end{aligned}
\right.
\end{equation}
Here we use $P(c'|q_i)$, the typicality of observing concept $c'$ when given $q_i$ to be the probability of starting from $q_i$ to $c$. 

Then, $\delta(c_i)$ will be:
\begin{equation}
\left\{
\begin{aligned}
\delta(c_i)&=1,&if c_i\in C_q^k\\
\delta(c_i)&=0,&if c_i\notin C_q^k
\end{aligned}
\right.
\end{equation} 

\paragraph*{The rationality of the model}
As for our running example, we can find that both \textsl{country} and \textsl{developing country} related to all entities in the query. However, it is obvious that \textsl{country} is a concept which is too general to be the underlying concept when given \textsl{\{China, India, Brazil\}}, and \textsl{developing country} may be a better one. It is easy to be observed that \textsl{country} is also a concept of \textsl{developing country}. That means there may exist some longer paths which will increase the expected steps starting from the query entity to the general concept. Thus, a concept not too general should be a concept which has a short expected distance to query entities. To avoid a concept too specific, the concept should have a short expected distance to every entities in the query set $q$. Thus, the optimization target should be to find $k$ concepts which can minimize $H(q|C_q^k)$.

Note that to make our solution efficient, two strategies can be used. First, $h(e|c)$ can be computed offline which will not increase the computation time when given queries to suggest entities. Second, since we only interested in concepts within a short hitting time, we may just ignore the concepts with a hitting time larger than a certain threshold of steps which can dramatically shorten the computation cost.

\subsection{$P(c_i|q)$ Computation}
In this part, we elaborate how to compute $P(c_i|q)$ which evaluate the extent of the concept underlying the query entities. 
\subsubsection{ Naive Bayes Model}

We first propose a Naive Bayes approach to compute $P(c_i|q)$.
According to the Bayes theorem, we have: 
\begin{equation}
P(c_i|q)=\frac{P(q|c_i)P(c_i)}{P(q)}\nop{\propto P(q|c_i)P(c_i)}\propto P(q|c_i)P(c_i)
\end{equation}
Since $P(q)$ is only dependent on the query, it can be ignored for the purpose of ranking.

In general,  a person's choices of two entities $e_i, e_j$ are logically independent with each other when given concept $c_i$. Thus, we can have an independence assumption that:
\begin{equation}
P(e_j,e_k|c_i)_{\forall e_j,e_k \in q}=P(e_j|c_i)P(e_k|c_i)
\end{equation} 
Then, we have:
\begin{equation}
P(c_i|q)\propto \prod_{e_j\in q}P(e_j|c_i)P(c_i)
\label{eq:bayes}
\end{equation} 

\nop{
Note that the objective $\argmax_e rel(q,e)$ is equivalent to 
\\ $\argmax_e \log r_q(e)$. The log ranking score of Eq.~\ref{eq:rank} is  
\begin{equation}
\log rel(q,e)=\sum_i[\log(P(e|c_i)\delta(c_i))+\log P(c_i|q)]
\end{equation}
where 
\begin{equation}
\log P(c_i|q)=\sum_{e_j\in q}\log P(e_j|c_i)P(c_i)
\end{equation} 
In the above equations, all probabilities are replaced with its corresponding log-likelihoods, which avoids underflow caused by multiplication of small values.}

Generally, there are relatively few concepts related to all of the query entities, therefore appropriate smoothing is necessary to avoid zero probabilities. 
To do this, we can assume that with probability $\lambda$, the user would choose the concept by its prior typicality. Thus, we can rewrite the Eq.\ref{eq:bayes}:
\\
$P(c_i|q)\propto$
\begin{equation}
P(c_i)\prod_{e_j\in q,n(e_j,c_i)>0}\lambda P(e_j|c_i)\prod_{e_j\in q, n(e_j,c_i)=0}(1-\lambda)P(e_j)
\end{equation}
\nop{
 we have
\begin{equation}
\log P(c_i|q)=\sum_{e_j\in q}\log[\lambda P(e_j|c_i)P(c_i)+(1-\lambda)P(c_i)]
\end{equation}}
where $P(e_j|c_i)$ can be computed by Eq.~\ref{eq:pec}, and $n(e_j,c_i)$ is the number of co-occurrences between $e_j$ and $c_i$ in Probase. The prior typicality of $P(c_i)$ and $P(e_j)$ can be computed by the following equation:
\begin{equation}
P(c_i)\propto n(c_i)
\label{eq:pc}
\end{equation} 
\begin{equation}
P(e_j)\propto n(e_j)
\end{equation}
where $n(c_i)$ and $n(e_j)$ is the number of occurrence of $c_i$ and $e_j$ in Probase.

Note that when using this model in the relative entropy model,  $P(c_i|q, e)$ depends on $P(q, e)$, which can not be ignored. Thus, the computation of $\frac{P(q,e)}{P(q)}$ in Eq.~\ref{eq:kld} can be simplified to be $P(e)$ by the independence assumption. 
\nop{Note that the above computation of $P(c_i|q)$ is appropriate for the probabilistic relevance model since the computation is independent of $e$. However, it is not appropriate for the relative entropy model because $P(c_i|q, e)$ cannot be computed like $P(c_i|q)$.  $P(c_i|q, e)$ depends on $P(q, e)$ and $e$ is a variable. Thus, the denominator in Eq.~\ref{eq:bayes} cannot be just ignored.}

\nop{
Hence, 
Here, we use the information in Probase to help us to continue to compute the probability of $p(q_i|c)$ and $p(c)$.
Thus, we have:
\begin{equation}
p(q_i|c)=\frac{n(q_i,c)}{n(c)}
\end{equation}
\begin{equation}
p(c)=\frac{n(c)}{\sum_{c\in C}\sum_{e\in E}n(c,e)}
\end{equation}
where $n(q_i,c)$ means the number of co-occurrence of the query entity $q_i$ and the concept $c$, and $n(c)$ means the number of the appearance of $c$ which equals to $\sum_{e\in E} n(c,e)$.

In next part, we will propose a method which performs better and more efficient when using our ranking model based on {\it concept distribution}.
}

\subsubsection{A Noisy-Or Model}

Alternatively, we can mimic a psychological process of
identifying an ideal concept, when query instances are presented one by one
to a human---As more query entities are given,
desirable concepts will amplify and eventually peak, as illustrated in Figure 2.
This observation implies that the signal indicating the right concept should be amplified when more entities are observed, and the single indicating the incorrect concept should be weakened. All these observations motivate us to use a Noisy-Or model to compute $P(c_i|q)$:
\begin{equation}
P(c_i|q)=1-\prod_{e_j\in q}(1-P(c_i|e_j))
\end{equation}  
where $P(c_i|e_j)=\frac{n(c_i,e_j)}{n(e_j)}$ can be computed by Probase data.

\captionsetup{margin=20pt,format=hang,justification=justified}
\begin{figure*}
\epsfig{file=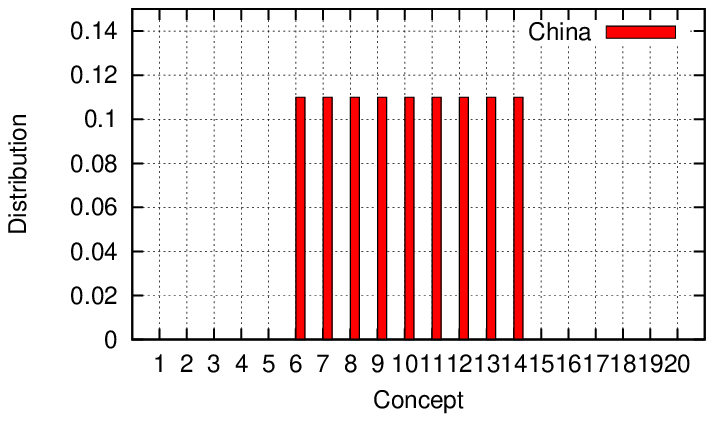,scale=0.9}
\epsfig{file=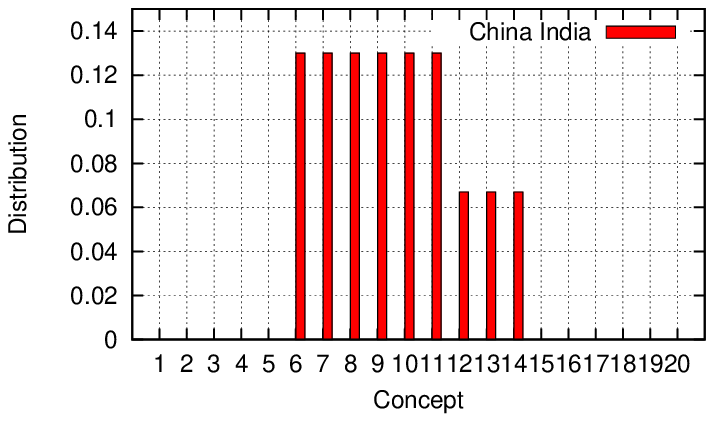,scale=0.9}
\epsfig{file=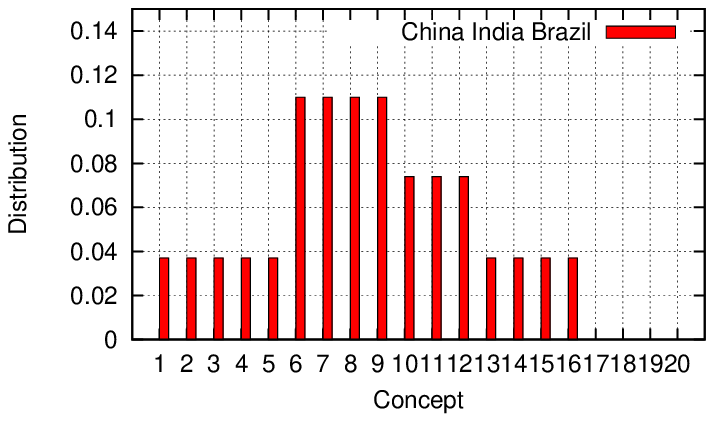,scale=0.9}
\epsfig{file=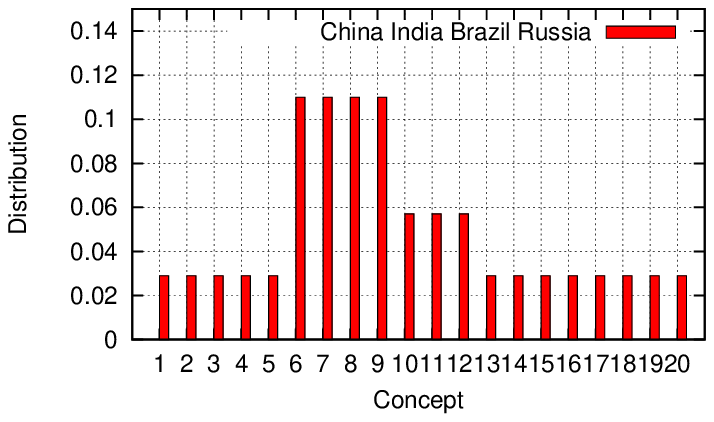,scale=0.9}
\centering
\caption{ Case study on the change of the concept distribution}
\label{pic:four}
\end{figure*}

Given $P(c_i|q)$, we can use it to compute $P(c_i|q,e)$ incrementally, which avoids the wasteful computation.
\begin{displaymath}
P(c_i|q,e)=1-\prod_{e_j\in q}(1-P(c_i|e_j))(1-P(c_i|e))
\end{displaymath}
\begin{displaymath}
=1-(1-P(c_i|q))(1-P(c_i|e))
\end{displaymath}
\begin{equation}
=P(c_i|q)+P(c_i|e)-P(c_i|q)P(c_i|e)
\end{equation}
where $P(c_i|q)$ is given and the other two parts can be computed by Probase.

\paragraph*{Rationality of Noisy-Or Model}
The Noisy-Or model reflects the following rationality: {\it the concept covering more query entities has a higher conditional probability}, which is consistent with our intuitions. We illustrate this by the following example. Still given \textsl{\{China, India, Brazil, Russia\}}, when we compute the query entities' distribution of their 20 concepts one by one, we can observe the change in Figure~\ref{pic:four}. It realized that the conditional probability became higher and concentrated to certain concepts. We also illustrate this in Example~\ref{exa:obs1}. 

\begin{example}[Rationality of Noisy-Or Model]
~\\
Given query entity set \textsl{\{China, India, Brazil\}}, if \textsl{\{China, India, Brazil\}} are all related to the concept \textsl{\{developing country, market\}}, it's obvious that these two concepts are both very important. The concept \textsl{Latin American country} is only related to the entity \textsl{Brazil}, so it will be less important to the concept \textsl{\{developing country, market\}}. Thus, the more query entities the concept is related to, the more important the concept is. Furthermore, the signal indicating the incorrect concept will be weakened.
\label{exa:obs1}
\end{example}

%% file: Experiment.tex
\section{Experiment Evaluation} \label{sec:exp}
In this section, we systematically evaluate the effectiveness of our models and solutions with the state-of-the-art approaches on two ground truth data sets. 

The first one is the data set presented in Wang and Cohen~\cite{wang2007} which are simple conceptual lists.
To highlight the advantage of our approaches, using fine-grained underlying concepts, we constructed a tougher data set to test our solutions. The second data set consists of lists that are deliberately selected from Wikipedia. These lists can only be described by complicated (fine-grained) concepts and might not have explicit simple concept names, and we call them complicated-yet-typical concepts. This data set is much more challenging because it is more likely to introduce false-positives because of the granularity of the concepts underlying the example entities. Then, we introduce our two data sets in detail and present the experimental results on each data set respectively.

We evaluate the effectiveness of our two models (PRM and REM) with two computation methods of $\delta(c_i)$ and $P(c_i|q)$ respectively, Penalizing popular concepts (PP for short), Granularity-aware approach (FG for short), Naive Bayes approach (BA for short) and Noisy-Or model (NO for short). Thus, we have overall 8 versions of our solution: PRM+PP+BA, PRM+FG+BA, PRM+PP+NO, PRM+FG+NO, REM+PP+BA, REM+FG+BA, REM+PP+NO, REM+FG+NO. We compared them with the baseline KNN method we proposed, the entire list completion proposed in~\cite{ER} (ER for short), the structure-based approach using properties of entities in~\cite{Example} (ESBA for short) and SEISA~\cite{Y}, one of the strongest baselines~\cite{wang2014}. Unluckily, since we do not have the web list data used in SEISA, we re-implemented it by replacing the web lists with the concepts and their entities in Probase.
 
\subsection{Simple Conceptual Lists}

\paragraph*{\bf Set up}

We use lists in English in the seal data set~\cite{wang2007} as our first test data set. We randomly choose two instances of each list as the query and evaluate the quality of the ranked result lists returned by different solutions.

\paragraph*{Competitors}
We compared our results with KNN, ESBA, SEISA and the results of ER reported in~\cite{ER}. Since ER has different versions, we report the performance results derived by the version that $P$ (or $R$, $F$) is best. We denote these versions by $ER_p$, $ER_r$, $ER_f$, respectively. 

Note that here we implement our approaches with granularity-aware approach with $k=200$ because in the first data set, concepts are all general concepts, thus we need to set a relatively large $k$, here 200, and we will study the effect when varying $k$ from different numbers.

\paragraph*{Metrics}
We use precision, recall and F-score as the metrics on our data set. Let the number of the entities in the ranked list be $n_R$ and the number of the entities in the seal list be $n_L$. Thus, precision $P$ is the number of correct entities that are in the ranked list divided by $n_R$. Recall $R$ is the number of the correct entities which appear in the ranked list, divided by $n_L$. F-scores are the harmonic mean of the recall and precision.

\paragraph*{\bf Results}
\paragraph*{Precision, Recall and F-score}
The comparison results in Table~\ref{ref:seal} show that almost all of our solutions consistently outperform the competitors on all the tested measures except the precision of REM+BA.  It sufficiently suggests that the conceptual taxonomy is beneficial for the entity retrieval task in most cases. 

The detailed comparisons reveal that PRM+FG+NO has the highest precision, and REM+FG+NO has the highest recall. When comparing our approaches with FG and PP, BA and NO respectively, we can find that the methods with FG are better than the ones with PP in most cases, and the methods with NO are always better than the ones with BA. It reveals that the Noisy-Or model computes the conditional concept distribution more accurately than the Naive Bayes Model, and very often the methods that directly penalize popular concepts may make some specific concepts greatly affect the results and make it more difficult to find related entities. 

\begin{table}
\centering
\normalsize
\caption{Results on the first dataset}
\begin{tabular}{ccccc} \hline
Method&P&R&F\\ \hline
KNN&0.13&0.230&0.166\\ \hline
ESBA&0.221&0.263&0.241\\ \hline
SEISA&0.240&0.286&0.259\\ \hline
PRM+PP+BA&0.257&0.363&0.317\\ \hline
PRM+FG+BA&0.294&0.449&0.355\\ \hline
PRM+PP+NO&0.310&0.409&0.359\\ \hline
PRM+FG+NO&\textbf{0.380}&0.537&\textbf{0.445}\\ \hline
REM+PP+BA&0.180&0.428&0.253\\ \hline
REM+FG+BA&0.196&0.422&0.268\\ \hline
REM+PP+NO&0.280&0.444&0.364\\ \hline
REM+FG+NO&0.362&\textbf{0.570}&0.443\\ \hline
$ER_p$&0.227&0.142&0.137\\ \hline
$ER_r$&0.121&0.236&0.136\\ \hline
$ER_f$&0.204&0.209&0.158\\ \hline
\label{ref:seal}
\end{tabular}
\end{table}

\begin{figure}
\centering
\epsfig{file=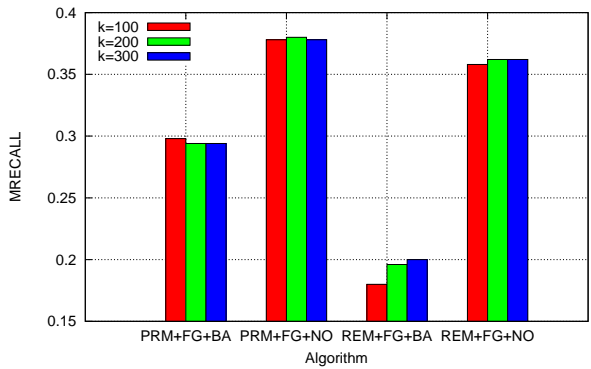,scale=0.8}
\caption{MAP of granularity-aware approach varying k }
\label{fig:pre}
\end{figure}

\begin{figure}
\centering
\epsfig{file=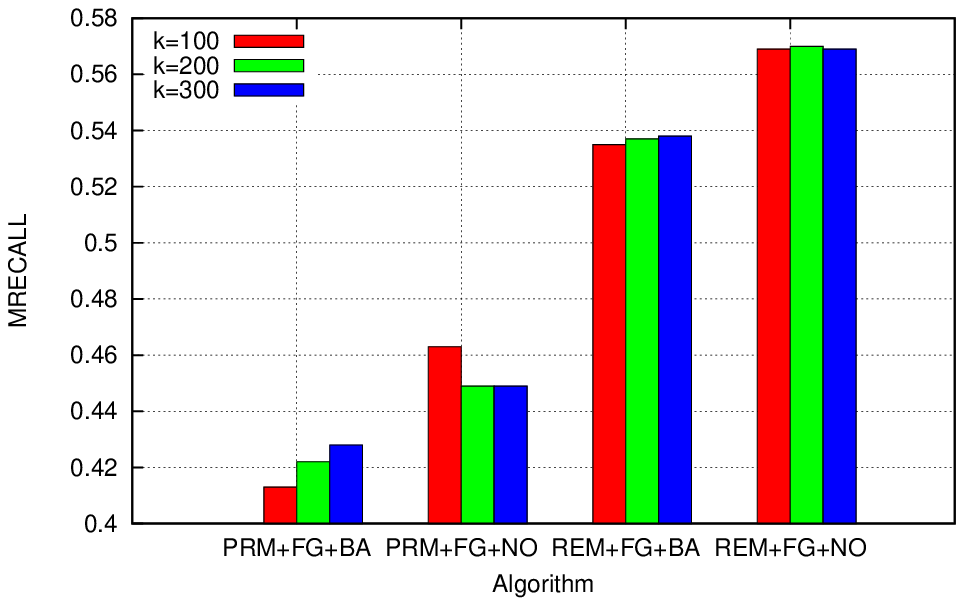, scale=0.8}
\caption{MRECALL of granularity-aware approach varing k}
\label{fig:rec}
\end{figure}

\paragraph*{Influence of \# concepts underlying the query}
In Figure~\ref{fig:pre} and~\ref{fig:rec}, we study the effect of the number of the concepts underlying the query examples on the results. As we have mentioned, the larger the number is, the more general concepts will be introduced into the concept distribution. However, here we can see that the number do not make a big difference. The reason is very likely to be the fact that the query examples generated by the lists of the seal data set are all from general concepts, thus the fine-grained concepts may not play an important role in generating the results.


\subsection{Fine-grained Conceptual Lists}

As we have noticed, the lists in seal data set have simple concept names, such as \textsl{\small us-presidents}. \textsl{\small Us presidents} is a specific and relatively simple concept, whose counterpart can be easily found in Probase. Thus, our task is reduced to the retrieval of entities from the concepts in Probase. This fact implies that our idea to use conceptual taxonomy is rational. On the other hand, this data set somewhat biases towards our solutions. Hence, it motivates us to construct a tougher data set, which sufficiently reveals the advantage of our models and solutions, specifically our approaches can find the entities of fine-grained conceptual lists with high precision. 

For this purpose, we deliberately select some complicated-yet-typical concepts from the Wikipedia articles. Most of these concepts have the name such as \textsl{\small Big N} and \textsl{\small Great N}. These article pages contain a list of entities that share the same specific concept, which usually requires a complicated description. For example, from the \textsl{\small Big 4 (tennis)} page in Wikipedia, we can get four entities \{\textsl{\small Roger Federer, Rafael Nadal, Novak Djokovic, Andy Murray}\}. 
The full description of the concept actually  is {\it the four most famous tennis players in the world nowadays}.  
We collected 112 such lists to construct the second data set. Some example lists as well as their complicated concept descriptions are shown in Table~\ref{tab:exalist}. 

\paragraph*{Competitors} We compared our solutions on the second data set with KNN, ESBA, SEISA and ER proposed in~\cite{ER} which implemented the entity ranking task with Bayesian Inference using Wikipedia data. To compare with it fairly, we re-implemented its method on Probase data. We used the concept in Probase as the word item (the hidden random variable $\theta$) in Wikipedia, and directly computed $P(e|\theta)$ (the conditional probability of entity $e$ given $\theta$), $P(\theta|D)$ (the conditional probability of $\theta$ given the example entities $D$), and $P(e)$ based on the probability in Probase. We denote this approach as ER+BS.


\begin{table*}
\centering
\normalsize
\caption{Fine-grained conceptual lists}
\begin{tabular}{|c|c|c|} \hline
Entity List&The full description of the concept\\ \hline
\tabincell{c}{Leonnado da vinci,\\ Raphael, Michelangelo}& \tabincell{c}{the three most famous art masters of the renaissance}\\ \hline
\tabincell{c}{Aaron Kwok, Jacky Cheung,\\ Leon Lai, Andy Lau}& the four most famous singers in hongkong\\ \hline
\tabincell{c}{Cats, Miss Saigon, Les Miserables,\\Phantom of the Opera}& the four most famous and classical musicals \\ \hline
\tabincell{c}{Agricultural Bank of China,\\ China Construction Bank, Bank of China,\\Industrial and Commercial Bank Of China}&the four biggest banks owned by chinese government\\ \hline
\end{tabular}
\label{tab:exalist}
\end{table*}

\paragraph*{Query construction}
Given the ground truth data, we use the following way to construct the query set for the evaluations:

For each list $L$ with $|L|$ instances, we use $\sigma(L)=\lceil \alpha|L| \rceil$ entities as the query entities. Ideally, we hope our solution can find the remaining $1-\alpha$ entities. In our experiment, we choose $\alpha$ as $\frac{1}{2}$, $\frac{2}{3}$, $\frac{3}{4}$, and we will also study its influence on the results.

\paragraph*{Metrics}
For each query, the answer entities should be ranked higher than other unrelated entities. Thus, on this data set, we use average NDCG to evaluate each query in $Q$, denoted as $mndcg$. Obviously, a larger $mndcg$ implies a better ranking. 

\paragraph*{\bf Results}

\paragraph*{mndcg}

\begin{figure}
\centering
\epsfig{file=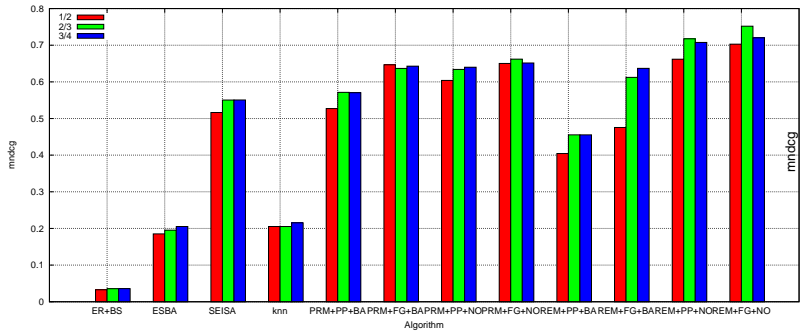,scale=0.5}
\caption{Performance of different solutions}
\label{exp:comp}
\end{figure}

The experimental results on the complicated conceptual lists are shown in Figure~\ref{exp:comp}. It is evident that except REM+BA, our approaches are all better than the baselines. We can find that REM+FG+NO has the highest mndcg even varying the number of the examples. Comparing the results in detail, it is obvious that FG+NO are always better than PP+BA, which reveals two conclusions that first, Noisy-Or model is more accurate in modeling the conditional probability when given some example entities. Second, granularity-aware approach is better in suggesting entities underlying fine-grained concepts. 

\begin{figure*}[!htb]
\centering
\epsfig{file=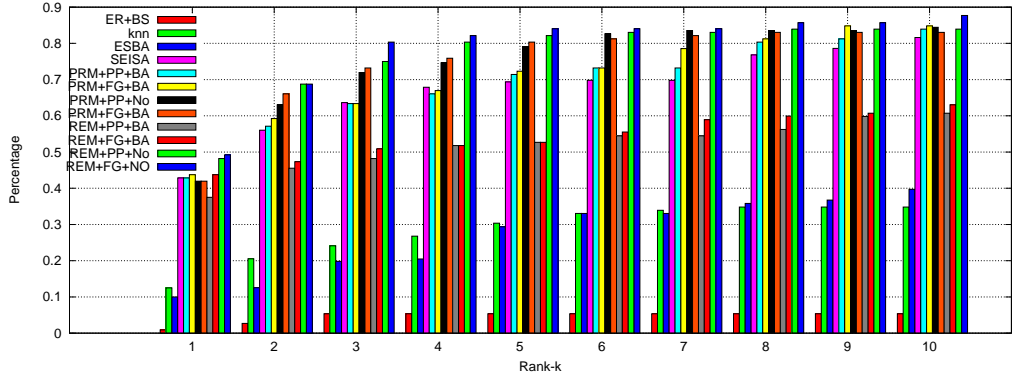,scale=0.72}
\caption{Precision@K of different methods}
\label{exp:comp2}
\end{figure*}

In Figure~\ref{exp:comp2}, we study the precision@k of different methods. We can find that our method REM+FG+NO has a relatively higher precision than other methods even when $k$ is small, and its precision can arrive at above 80\% when $k$ is 3. Furthermore, PRM+NO models also perform well when $k$ is small which reveals the rationality of our models. 

\paragraph*{Influence of \# concepts underlying the query }
We study the choice of the number of the concepts underlying the query examples. We vary k from 20, 50, 100 to 200 to observe the results. We can find that when we choose 50 or 100, the $mndcg$ may be higher than we choose 20 or 200. It is reasonable because 20 fine-grained concepts may be a little too specific to find accurate related entities, while 200 concepts may introduce more general concepts leading to incorporating many false positive entities. Therefore, k, the number reflecting the granularity should be carefully chosen.

\begin{table*}
\centering
\normalsize
\caption{Top-5 results of four cases}
\begin{tabular}{|c|c|c|c|} \hline
\multicolumn{2}{|c|}{\textbf{Case 1}}&\multicolumn{2}{c|}{\textbf{Case 2}}\\ \hline
\multicolumn{2}{|c|}{\tabincell{c}{\textbf{Query:}~alibaba, tencent}}&\multicolumn{2}{c|}{\tabincell{c}{\textbf{Query:}~\tabincell{c}{agriculture bank of china, china construction bank}}}\\ \hline
\multicolumn{2}{|c|}{\textbf{Possible Answer:}~baidu}&\multicolumn{2}{c|}{\textbf{Possible Answer:}~\tabincell{c}{bank of china, industrial and commercial bank of china}}\\ \hline
REM+FG+NO&SEISA&REM+FG+NO&SEISA\\ \hline
baidu&ebay&icbc&bank of china\\ \hline
neusoft group&baidu&bank of china&bank\\ \hline
alibaba.com & amazon & china development bank&bank of america\\ \hline
lenovo group&facebook&bank of communications&commercial bank of china\\ \hline
byd&global source&china international capital corporation limited &industrial\\ \hline
\end{tabular}
\label{tab:big1}
\end{table*}

\begin{table*}
\centering
\normalsize
\caption{Top-5 results of four cases}
\begin{tabular}{|c|c|c|c|} \hline
\multicolumn{2}{|c|}{\textbf{Case 3}}& \multicolumn{2}{c|}{\textbf{Case 4}}\\ \hline
\multicolumn{2}{|c|}{\tabincell{c}{\textbf{Query:}~\tabincell{c}{aaron kwok, jacky cheung}}}&\multicolumn{2}{c|}{\tabincell{c}{\textbf{Query:}~\tabincell{c}{kmpg, pricewaterhousecoopers} }}\\ \hline
\multicolumn{2}{|c|}{\textbf{Possible Answer:}~leon lai, andy lau}& \multicolumn{2}{c|}{\textbf{Possible Answer:}~ernst, deloitee}\\ \hline
REM+FG+NO&SEISA&REM+FG+NO&SEISA\\ \hline
andy lau	&madonna&ernst \& young&ibm\\ \hline
leon lai&kanye west&deloitte&accenture\\ \hline
edison chen&bette midler&ernst&fidelity investments\\ \hline
george guilder&beyonce&microsoft&schwab\\ \hline
george shea&aretha franklin&pwc&acs\\ \hline
\end{tabular}
\label{tab:big2}
\end{table*}

\begin{table*}
\centering
\normalsize{
\caption{Top-5 concepts}
\begin{tabular}{|c|c|c|c|} \hline
\textbf{Case 1}&\textbf{Case 2}&\textbf{Case 3} & \textbf{Case 4}\\\hline
chinese internet giant& chinese bank&\tabincell{c}{popular canto-pop \\entertainer}&accounting firm\\ \hline
\tabincell{c}{shanghai-chinese \\internet giant}&mainland bank&hot hongkong singer&global consultant\\ \hline
B2b service provider &chinese government bank&hongkong artists&\tabincell{c}{international\\ accounting firm}\\ \hline
chinese consumption stock&chinese leader&\tabincell{c}{universal's other\\ frontline artist}&big accounting firm\\ \hline
\tabincell{c}{chinese social \\networking site}&bank&spokesperson&\tabincell{c}{global business \\consultancy firm}\\ \hline
\end{tabular}
\label{tab:concept}
}
\end{table*}

\subsection{Case studies}

In this subsection, we give case studies to show the rationality of our models and solutions.
 
Table~\ref{tab:big1} and~\ref{tab:big2} presents studies on some cases from our second data set. The query examples are in the second row of the tables, and the entities in the third row are answer entities which complete the complicated-yet-typical concepts where the example entities come from. They are possible entities which people may want the computer to return when given the query entities. Of course, some other entities are also very likely to be associated when given these entities, but here we consider that the result lists which contain these possible answers have a higher quality. 

\begin{figure}[!htb]
\centering
\epsfig{file=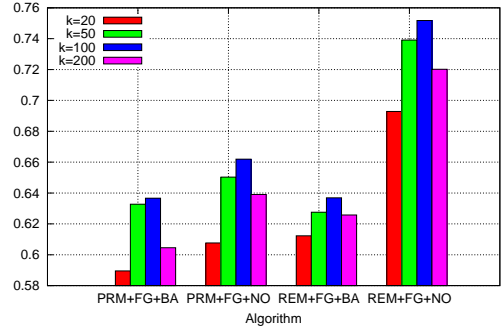,scale=0.72}
\caption{Varying k of granularity-aware method}
\end{figure}
We present the top-5 entities of the result lists. Because of the space limitation, we show the results of the best competitor and the best approach of ours. In all of the cases, the possible answers appear in the top of the lists of our results. While the baseline tend to introduce false positive entities, and give them higher ranks. Concretely, in Case 1, \textsl{alibaba} and \textsl{tencent} are Chinese internet giants, while SEISA incorporated a lot of companies in the U.S. It is very likely that SEISA used the concept \textsl{company} leading to the false positive entities. The same situation also happened in Case 2, and Case 3. In Case 4, SEISA seemed to include some unrelated entities.

To testify the contribution of the fine-grained concepts, in Figure~\ref{tab:concept}, we show the latent results of entity suggestion. Here, we show top-5 concepts mined by our method. We can see that most of our concepts are related and not too general or specific.

The above results sufficiently show that the biased selection of concepts is critical for the effectiveness of entity inference.

\nop{
\begin{table}
\centering
\caption{Case 2}
\begin{tabular}{|c|c|c|c|} \hline
\multicolumn{}{|c|}{Query:Pricewaterhousecoopers,KPMG,Ernst}\\ \hline
PRM+BA&PRM+NO&REM+NO&knn \\ \hline
Young&Young&IBM&Bain\\ \hline
\tabincell{c}{Ernst \\ \& Young}& \yh{Deloitte}&\yh{Deloitte}&\yh{Deloitte}\\ \hline
\yh{Deloitte}&\tabincell{c}{Ernst \\ \& Young}&\tabincell{c}{accenture}&\tabincell{c}{Ernst \\ \& Young}\\ \hline
IBM&IBM&Microsoft&\tabincell{c}{Arthur \\Andersen}\\ \hline
accenture&accenture&\tabincell{c}{Ernst \\ \& Young}&accenture\\ \hline
\end{tabular}
\end{table}

\begin{table}
\centering
\caption{Case 2}
\begin{tabular}{|c|c|c|c|} \hline
\multicolumn{4}{|c|}{Query:Zeus,Hades}\\ \hline
PRM+BA&PRM+NO&REM+NO&knn \\ \hline
\yh{Poseidon}&Apollo&\yh{Poseidon}&Apollo\\ \hline
Apollo&\yh{Poseidon}&Apollo&\yh{Poseidon}\\ \hline
Athena&Aphrodite&Aphrodite&Trojan\\ \hline
Odin&Athena&Athena&Vundo\\ \hline
Apache&Hermes&Herms&Apache\\ \hline
\end{tabular}
\end{table}
}